# Energy Relaxation Time between Macroscopic Quantum Levels in a Superconducting Persistent Current Qubit

Yang Yu, D. Nakada, Janice C. Lee, Bhuwan Singh, D. S. Crankshaw, and T. P. Orlando

Massachusetts Institute of Technology, Cambridge, Massachusetts 02139, USA

William D. Oliver and Karl K. Berggren

MIT Lincoln Laboratory, Lexington, Massachusetts 02420, USA

**Abstract**

We measured the intrawell energy relaxation time $\tau_d$ between macroscopic quantum levels in the double well potential of a Nb persistent-current qubit. Interwell population transitions were generated by irradiating the qubit with microwaves. Zero population in the initial well was then observed due to a multi-level decay process in which the initial population relaxed to the lower energy levels during transitions. The qubit's decoherence time, determined from $\tau_d$, is longer than 20 μs, holding the promise of building a quantum computer with Nb-based superconducting qubits.

PACS numbers: 03.67.Lx, 03.65.Yz, 85.25.Cp, 85.25.Dq

Recent successes in superconducting qubits (SQ) have enhanced the feasibility of implementing quantum computing (QC) with Josephson devices [1-9]. Rabi oscillations, which are a preliminary requirement of QC, have been reported in charge, phase, and flux qubits [3-8]. However, the systematic experimental investigation of the decoherence, which is a key issue of SQ, is sparse so far due to the challenge of the time resolution of the measurement. Although long decoherence times have been shown in some special configurations [4,5,8,10], the limiting source of decoherence in the SQ remains unidentified. On the other hand, for SQ the decoherence time, including energy and phase relaxation time, is predicted to be proportional to the level of dissipation, which results from the coupling between the qubits and the environment [11-13]. Therefore, determination of the dissipation is extremely useful in designing new qubits with various materials because it will show whether the dissipation of the qubits is low enough to make error-tolerant computation possible. Previous methods to determine the dissipation of devices are either applicable at relatively high temperatures [10], or rely on indirect measurements of switching probabilities [14,15]. In addition, all long decoherence times (~ 1 µs) reported were obtained in NbN and Al Josephson devices [4,5,8,10]. It is important to know whether a promising decoherence time can be achieved in Nb-based Josephson devices, which has a more mature fabrication capability. In this Letter, we present time-resolved measurements of the intrawell relaxation time $\tau_d$ in a Nb persistent-current (PC) qubit. It was found that $\tau_d \simeq 24$ µs, indicating a decoherence time longer than 20 µs. This long decoherence time demonstrates a strong potential for QC employing Nb-based SQ.

A PC qubit is a superconducting loop broken by three under-damped Josephson junctions (JJs) (Fig. 1 (a)). Two JJs are designed to have the same critical current, and the third one is $\alpha$ times smaller. For $0.5 < \alpha < 1$ and with an externally applied magnetic field close to a half flux quantum, $\Phi_0/2$, the system behaves as a particle moving in a double well potential (Fig.1 (b)), with the classical states in each well corresponding to macroscopic persistent currents of opposite sign [16]. The potential can be tilted back and forth by changing the frustration $f_q$, which is the magnetic flux threading the loop in units of $\Phi_0$. At low temperature and weak damping, the dynamics of the system are described by quantum mechanics, and the particle occupies quantized energy levels in each well. The two classical states are then coupled via quantum tunneling through the barrier between the wells. In addition, the system can interact quantum mechanically with a monochromatic electromagnetic (microwave) field, and microwaves with frequency matching the energy level spacing can generate transitions between the two macroscopic quantum states, namely, photon induced transitions (PIT) [2,8].

The samples used in this study were fabricated at MIT Lincoln Laboratory in a Nb trilayer process [17]. The critical temperature of the JJs was $T_c \simeq 9.1$ K. The PC qubit area is 16×16 µm$^2$, with a self-inductance of $L_q \simeq 30$ pH. The critical current density was $J_c \sim 370$ A/cm$^2$. The critical currents of the large and small JJs in the qubit, determined from thermal activation studies [18], were $I_c \simeq 1.2$ µA and 0.75 µA respectively ($\alpha \simeq 0.63$). The qubit energy level structure calculated using these parameters is shown in Fig. 1 (c). The readout dc SQUID, which surrounds the qubit, consists of two JJs with equal critical current $I_{c0} \simeq 6$ µA. Both JJs were shunted with a 1-pF capacitor to lower the resonant frequency of the SQUID. The SQUID is $20\times20$ $\mu m^2$ in area, with a self-inductance of $L_{SQ} \simeq 60$ pH. The mutual inductance between the qubit and dc SQUID is $M \simeq 25$ pH. The inductances and $I_{c0}$ were determined from SQUID transfer function measurements and are consistent with the estimated values from the fabrication process parameters. The persistent current in the qubit loop will generate an additional magnetic flux of $\sim 3$ m$\Phi_0$ in the SQUID, resulting in a 0.1 µA change in the switching current $I_{sw}$ of the SQUID that can be easily detected at $T < 50$ mK. The sample was mounted on a chip carrier that was enclosed in an oxygen-free-copper sample cell, which was thermally anchored to the mixing chamber of a dilution refrigerator. The devices were magnetically shielded by four cryoperm-10 cylinders surrounding the inner vacuum can. All electrical leads that connected the SQUID to room temperature electronics were carefully filtered by EMI filters (at 300 K), RC filters (at 1.6 K) and copper powder filters (at 15 mK). Microwaves were injected to the qubit via a separate semirigid cryogenic coaxial cable with 20 dB attenuators at the 1 K pot and the mixing chamber. Battery-powered low-noise pre-amplifiers were used for all measurements. The switching current distribution measurement of a test junction indicated that there was no significant extrinsic noise in our system.

The spectroscopy of the qubit energy levels was examined by using microwave pulses to produce PITs. For each measurement trial (Fig. 2), we first prepared the qubit in state $|1\rangle$ by tilting the potential (i.e., applying frustration) to where the system has a single well and waiting a sufficiently long time. After the qubit relaxed to its ground state, the potential was tilted back to the frustration where it was to be measured. At low temperatures, the qubit will have a finite probability of remaining in $|1\rangle$, which is effectively metastable on the timescales considered in this paper. We then applied microwaves with duration time $t_{pul}$, inducing transitions between states $|1\rangle$ and $|0\rangle$. After the microwaves were shut off, the bias current of the SQUID was ramped through values slightly higher than the critical current $I_{c0}$. The qubit state ($|0\rangle$ or $|1\rangle$) was then read out from the current at which the SQUID switched to a finite voltage state (**0** or **1** in Fig. 2 (d)). For a fixed frustration, this procedure was repeated more than $10^3$ times to minimize the statistical error. A histogram of switching currents clearly shows the probability distribution of the qubit state occupation. Shown in Fig. 3 are contour plots of the switching-current histograms by scanning the frustration at $T = 15$ mK. Each vertical slice is a switching current histogram, and color bars represent the number of switching events (or probability). A split of the switching-current distribution, caused by the opposite persistent current of the qubit, was observed at $f_q \sim 0.485\Phi_0$. The lower branch represents the qubit in the $|0\rangle$ state, and the higher branch represents the qubit in the $|1\rangle$ state. The substantial population in state $|1\rangle$ demonstrates that we successfully prepared the qubit in $|1\rangle$, because, near $f_q \sim 0.485\Phi_0$, the qubit had a much higher ground state energy in $|1\rangle$ than that in $|0\rangle$. However, the energy barrier and width relative to the lowest energy level of state $|1\rangle$, denoted as $|1\rangle_0$, were small enough so that the qubit had a large probability of tunneling to $|0\rangle$. The tip of the higher branch marked a fixed frustration point $f_q \approx 0.484\Phi_0$, below which it was impossible for the qubit to stay in $|1\rangle$, because the potential becomes a single well of $|0\rangle$ state. Microwaves, with frequencies matched the energy difference between the ground level of $|1\rangle$ and one of the levels of $|0\rangle$, were used to generate transitions between $|1\rangle$ and $|0\rangle$. The most striking feature of the contour plots is that a population "gap" (i.e., zero population) in the $|1\rangle$ branch was created by the microwaves (Fig. 3 (b) to (d)). With increasing microwave frequency, the gap moved away from the tip, as expected from the energy level structure (Fig.1 (c)). The quantitative agreement between the gap position and the energy level structure confirmed that the gap resulted from the microwave PIT between the two macroscopic quantum states $|1\rangle_0$ and $|0\rangle_3$ (the third excited energy level of the state $|0\rangle$). We believe that the PIT here was an incoherent process, because the microwave pulse duration was 600 μs, which is much longer than the estimated decoherence time (0.1~100 μs) [2,8,16]. Additionally, no periodic variation of population with varying pulse duration was observed. Nevertheless, observing a gap is unexpected for an incoherent transition, since the population on the lower level ($|1\rangle_0$) should always be larger than 0.5 for a simple two level system [19]. In order to address this phenomenon, a multi-level pump-decaying model is introduced.

For simplicity we considered only three levels, the initial state $|1\rangle_0$, the $|0\rangle_3$ state to which radiation induces a transition, and the state $|0\rangle_2$ to which the population of $|0\rangle_3$ decays. The state $|0\rangle_3$ decays to $|0\rangle_2$, $|0\rangle_1$, and $|0\rangle_0$, but for ease of calculation we collectively label these states as $|0\rangle_2$ with an overall effective intrawell decay rate $\gamma_d \equiv 1/\tau_d$. The temporal evolution of the three-level system under microwave irradiation is thereby described by the following three coupled rate equations

$$\frac{dP_{10}}{dt} = -\gamma_1 P_{10} + (\gamma_1 + \gamma_2) P_{03},\tag{1}$$

$$\frac{dP_{03}}{dt} = \gamma_1 P_{10} - (\gamma_1 + \gamma_2) P_{03} - \gamma_d P_{03},\tag{2}$$

$$\frac{dP_{02}}{dt} = \gamma_d P_{03},\tag{3}$$

where $P_{10}$, $P_{03}$, and $P_{02}$ are the occupation probabilities of level $|1\rangle_0$, $|0\rangle_3$, and $|0\rangle_2$ respectively. $\gamma_1$ is the stimulated transition rate between $|1\rangle_0$ and $|0\rangle_3$, and $\gamma_2$ is the spontaneous relaxation rate from $|0\rangle_3$ to $|1\rangle_0$. Generally, for a given system, $\gamma_1$ is proportional to the microwave power $P_{rf}$, and $\gamma_2$ can be considered to be a constant [19]. For the initial condition $P_{10}(0) = 1$, while $P_{03}(0) = P_{02}(0) = 0$, eqs. (1)-(3) can then be solved analytically. For $\gamma_1 \gtrsim \gamma_d$, which is satisfied in our experiment, the probability of finding the qubit remaining in the state $|1\rangle_0$ at $t > 1/(2\gamma_1 + \gamma_2 + \gamma_d)$ is given by

$$P_{10}(t) \approx a_1 e^{-t/\tau'},\tag{4}$$

where $a_1$ depends weakly on the microwave power and can be considered as a constant in the relevant time scale,

$$\tau' \simeq (2 + \gamma_2/\gamma_1)\tau_d = (2 + \gamma_2/AP_{rf})\tau_d,\tag{5}$$

and $A$ is the coupling constant between the microwave source and the qubit. The physical picture of the three-level pump-decaying process is that microwaves populate the highest level with a population $P_{03} \propto 1/(2 + \gamma_2/\gamma_1)$, which decays to the lowest level with a rate $\gamma_d$. Therefore, the effective decay rate of the population of the initial state is given by eq. (5), and with $t$ sufficient long, $P_{10}(t) \to 0$; this agrees with the experimental observations.

A significant impact of Eqs. (4) and (5) is that $\tau_d$ can be determined by measuring $P_1(t)$. Because the switching current of $|0\rangle$ is smaller than that of $|1\rangle$, pumping the system from state $|1\rangle$ to state $|0\rangle$ will generate a dip in the switching current average as a function of frustration, and the dip amplitude is proportional to $1-P_{10}$. Fig. 4 shows the dip amplitude as a function of the microwave irradiation time, $t_{pul}$. The nominal power of the microwave source was $P_{rf} = 31.3$ μW. The time constant $\tau'$, obtained from a best fit, is $130 \pm 20$ μs. It should be emphasized that $\tau'$ is not equal to $\tau_d$, but, rather, depends on $\gamma_2/\gamma_1$. For large $P_{rf}$, (i.e., $\gamma_1 \gg \gamma_2$), $\tau'$ will saturate to $2\tau_d$. For $\gamma_1 \sim \gamma_2$, we are able to determine $\tau_d$ by measuring the microwave power dependence of $\tau'$. Shown in Fig. 5 is $\tau'$ measured at various microwave powers. $\tau'$ saturates at about 50 μs for $P_{rf} > 0.2$ mW. By adjusting $\gamma_2/A$ and $\tau_d$ as fitting parameters, we obtained $\tau_d \simeq 24.3 \pm 2.7$ μs from a best fit to Eq.(5), which is of the same order of magnitude as that reported energy relaxation times in NbN and Al-based qubits. Note that $\gamma_2$ is another important parameter which determines interwell energy relaxation. Unfortunately, we could not directly extract $\gamma_2$ from the fitting, because we do not know the coupling constant $A$. Future experiments in which microwave coupling is independently characterized should lead to the extraction of $\gamma_2$.

The primary effect of the environmental dissipation on the intrawell dynamics of the PC qubits is that, at low temperature ($k_B T \ll$ level spacing), the width of an excited level with energy $E_n$ is given approximately by $\gamma_d \simeq E_n/Q$, where $Q$ is the quality factor of classical small oscillation in the potential well [20,21]. From $\tau_d$ we determined $Q \sim$ $5\times10^5$, close to the value obtained from thermal activation measurements at intermediate temperatures 0.3 ~ 1.2 K [18]. Note that the theoretical subgap resistance depends on the temperature as ~ $e^{\Delta_s/k_BT}$ [22], where $\Delta_s$ is the superconducting gap voltage. The temperature independence of $Q$ suggests the presence of additional environmental sources of dissipation [16].

This long intrawell relaxation time is important for experiments in quantum computing in two ways. First, the lower two energy levels in the qubit, $|0\rangle_0$ and $|0\rangle_1$ could be used as the two qubits states, with the third state $|0\rangle_3$ used as the read-out state. Because our PC qubit had no leads directly connected to it and the magnetic coupling circuit is optimally designed to lessen the effects of the electromagnetic environment, the PC qubit is much less influenced by this environment than are other similar single-junction schemes [5,6,9]. Second, if we assume that the environment can be modeled as an ohmic bath, as in the spin-boson model, then we can estimate what the decoherence times of a flux qubit when the qubit states are interwell states of opposite circulating current [2,8,16]. The energy relaxation and phase decoherence times are given in the spin-boson model for an ohmic environment by [11,12]

$$\tau_{relax}^{-1} \simeq \pi\alpha_L \sin^2\eta\Delta E/\hbar, \qquad (6)$$
$$\tau_{\varphi}^{-1} = \tau_{relax}^{-1}/2 + 2\pi\alpha_L k_B T \cos^2\eta/\hbar, \qquad (7)$$

where $\Delta E$ is the energy difference between levels in opposite wells, $\eta \approx \text{tg}^{-1}(\Delta/\Delta E)$ is the mixing angle, $\Delta$ is the tunneling amplitude between the wells, and $\alpha_L \sim 1/Q$ is the quantum damping parameter [21] which we estimate to the our measured value. For our Nb PC qubit operating with opposite circulating currents states, a conservative estimate gives $\tau_{relax} \gtrsim 28$ μs and $\tau_\varphi \gtrsim 25$ μs at 15 mK. For a typical Rabi frequency $\Omega = 1$ GHz, we obtained a quantum quality factor $> 10^4$, larger than the basic requirement of the error-tolerant QC. Considering their attractiveness from the point of view of robust and well developed fabrication methods, Nb-based superconducting qubits are very promising for realizing scalable quantum computer.

In summary, we directly measured the intrawell relaxation time of a Nb-based PC qubit by generating PITs between macroscopically distinct quantum states. A multi-level decay process was observed with an intra well relaxation time of about 25 microseconds, with a $Q$ factor of greater than $10^5$, showing that these intrawell levels form are well isolated from the electromagnetic environment and would make a good qubit. Likewise, these measurements indicate that the flux qubits operating between wells would also have sufficient decoherence times, demonstrating good prospects for well-fabricated Nb junctions, with its more mature technology, to be used as superconducting qubits.

We thank K. Segall, B. Cord, D. Berns, and A. Clough for technical assistance and S. Valenzuela, M. Tinkham, L. Levitov, and M. Vavilov for helpful discussions. This work was supported by AFOSR Grant No. F49620-01-1-0457 under the Department of Defense University Research Initiative on Nanotechnology (DURINT). The work at Lincoln Laboratory was sponsored by the Department of Defense under the Air Force, Contract No. F19628-00-C-0002.

**Figure Captions**

Fig. 1. (a) Schematic of the PC qubit surrounded by a readout dc SQUID. (b) Schematic of the qubit's double-well potential with energy levels for an applied frustration close to $0.485\Phi_0$. Microwaves pump the qubit from the lowest level of $|1\rangle$ ($|1\rangle_0$) to the third excited level of $|0\rangle$ ($|0\rangle_3$), then decay to the second excited level of $|0\rangle$ ($|0\rangle_2$) with a rate $\gamma_d$. (c) Calculated energy-level diagram of the PC qubit using qubit parameters determined from independent measurements. The dashed circle marked where the PIT occurred.

Fig. 2. (a) Time profiles of bias frustration, (b) microwave amplitude, (c) SQUID bias current, (d) and SQUID voltage for one measurement trial. **0** and **1** indicate that the qubit states ($|0\rangle$ or $|1\rangle$) cause different switching currents.

Fig. 3. (a) Contour plots of the switching current distribution without microwaves, with microwaves at (b) $\nu = 6.77$ GHz, (c) 7.9 GHz, and (d) 9.66 GHz. The horizontal axis of each plot corresponds to frustration from $0.4835\Phi_0$ to $0.4856\Phi_0$. The arrows mark the gap moving away from a reference frustration.

Fig. 4. The amplitude of the microwave resonant dip as a function of microwave duration time. The microwave frequency $\nu = 9.66$ GHz, and nominal power $P_{rf} = 31.3$ μW. The solid squares are experimental data and the line is a best fit to an exponential decay. The inset shows the resonant dips at $t_{pul} = 0.2, 0.5, 0.8$ and 1 ms, from the top to the bottom.

Fig. 5. $\tau$ vs. microwave amplitude for $\nu = 9.66$ GHz. The solid line is a best fit to Eq. (5).

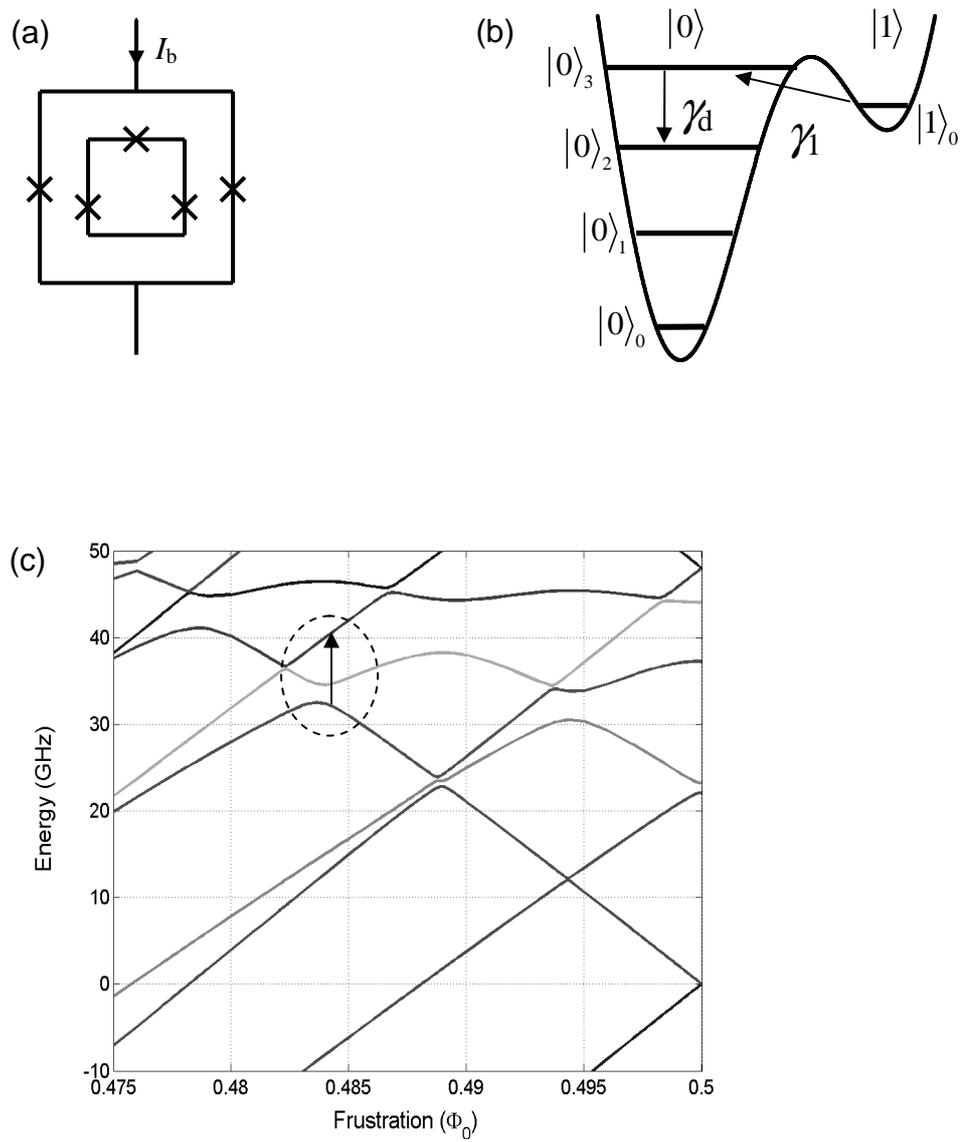

Fig. 1. Yang Yu *et al.*

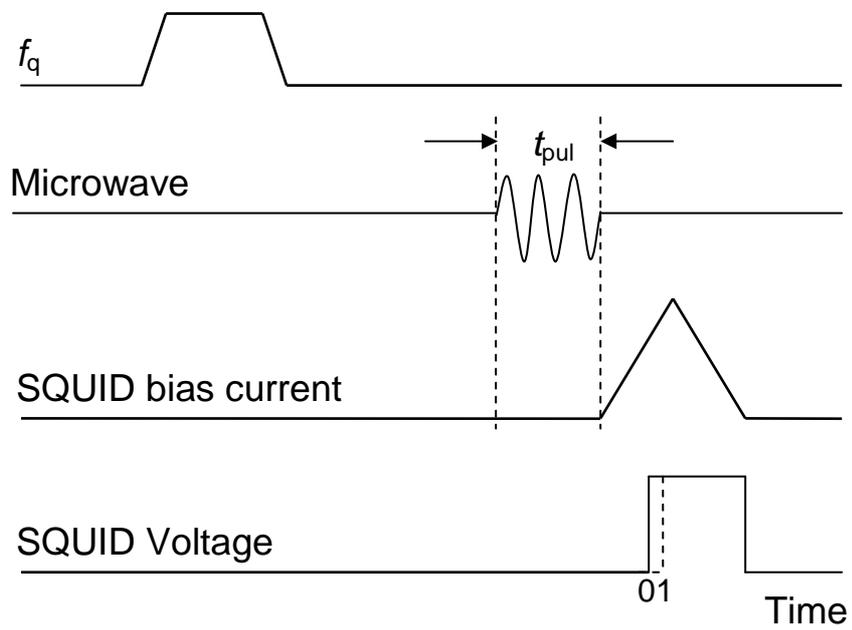

Fig. 2. Yang Yu *et al.*

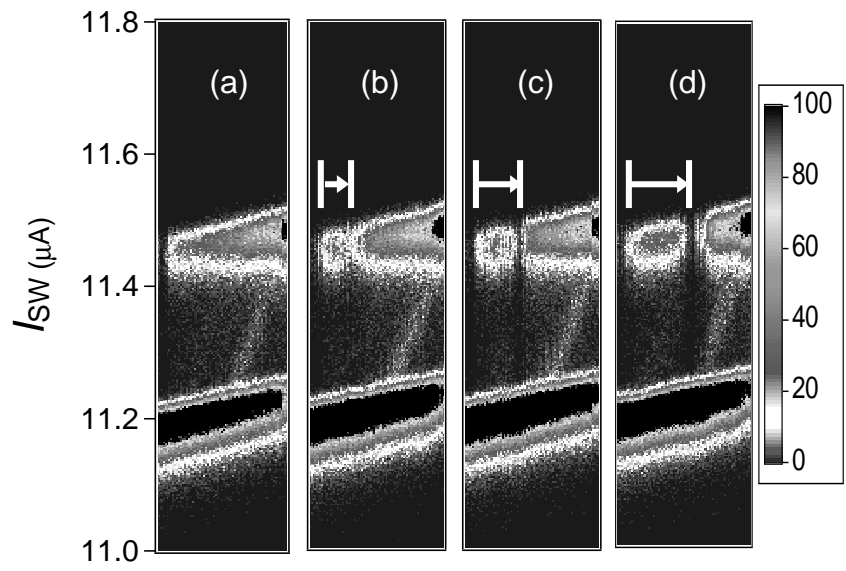

Fig. 3.   Yang Yu *et al.*

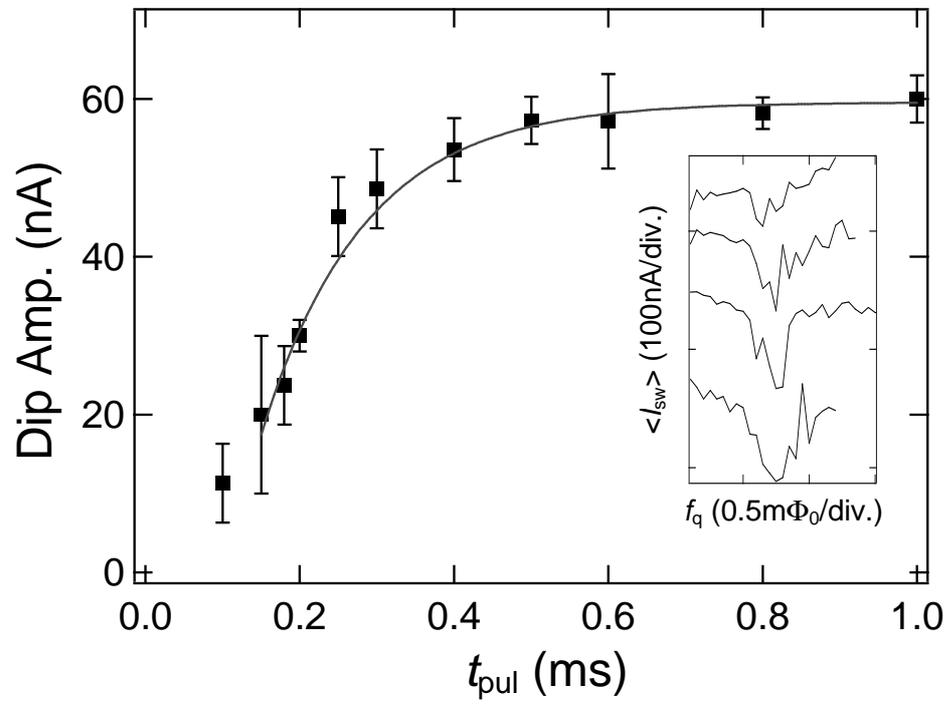

Fig. 4. Yang Yu *et al.*

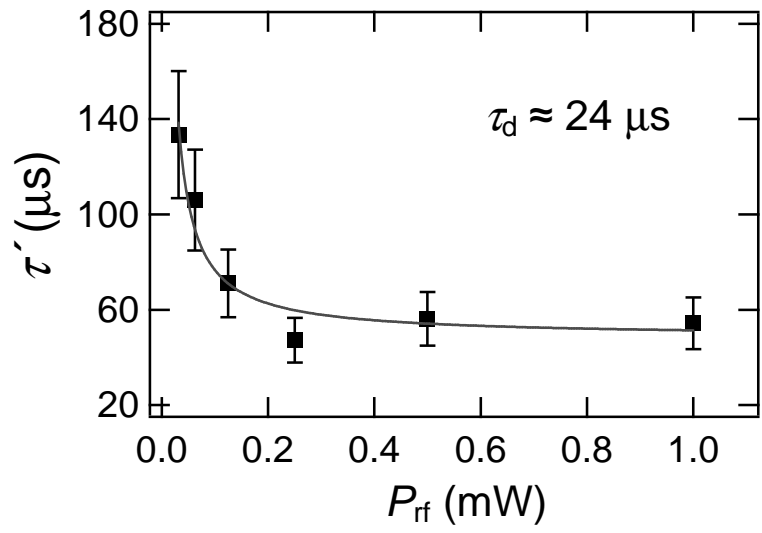

Fig. 5.   Yang Yu *et al.*